# REVISED, Accepted by journal 'Epidemiology', 23 June 2020

REVIEW ARTICLE

# The test-negative design with additional population controls: a practical approach to rapidly obtain information on the causes of the SARS-CoV-2 epidemic

Jan P Vandenbroucke[1,2,3], Elizabeth B Brickley[2], Christina M.J.E. Vandenbroucke-Grauls[4], Neil Pearce[2]

(1) Department of Clinical Epidemiology, Leiden University Medical Center, The Netherlands; (2) Departments of Medical Statistics, Non-communicable Disease Epidemiology and Infectious Disease Epidemiology, London School of Hygiene and Tropical Medicine, London, UK; (3) Department of Clinical Epidemiology, Aarhus University, Denmark; (4) Department of Medical Microbiology and Infection Prevention, Amsterdam UMC, The Netherlands

**Corresponding Author**: Jan P Vandenbroucke, Leiden University Medical Center, Dept. Clinical Epidemiology, PO Box 9600, 2300 RC Leiden, The Netherlands, E-mail: J.P.Vandenbroucke@lumc.nl

**Running head**: Test-negative design combined with population controls

The authors have no conflict of interest

No specific sources of financial support, other than salaries by affiliations

**Word count**: approximately 4600, two figures

**Acknowledgments:** We acknowledge our debt to comments on a previous version of this paper by Ivo Foppa, Mike Jackson, Daniel Westreich, Sheena Sullivan, Martie van Tongeren, several Twitter commentators (David Simons, Mark D Simmons), and Marc Lipsitch; for comments on Appendix B we thank Tim Lash and Avnika Amin. All opinions as well as remaining inadequacies are our own.

# Abstract

Testing of symptomatic persons for infection with SARS-CoV-2 is occurring worldwide. We propose two types of case-control studies that can be carried out jointly in test-settings for symptomatic persons. The first, the test-negative case-control design (TND) is the easiest to implement; it only demands collecting information about potential risk factors for COVID-19 from the tested symptomatic persons. The second, standard case-control studies with population controls, requires the collection of data on one or more population controls for each person who is tested in the test facilities, so that test-positives and test-negatives can each be compared with population controls. The TND will detect differences in risk factors between symptomatic persons who have COVID-19 (test-positives) and those who have other respiratory infections (test-negatives). However, risk factors with effect sizes of equal magnitude for both COVID-19 and other respiratory infections will not be identified by the TND. Therefore, we discuss how to add population controls to compare with the test-positives and the test-negatives, yielding two additional case-control studies. We describe two options for population control groups: one composed of accompanying persons to the test facilities, the other drawn from existing country-wide health care databases. We also describe other possibilities for population controls. Combining the TND with population controls yields a triangulation approach that distinguishes between exposures that are risk factors for both COVID-19 and other respiratory infections, and exposures that are risk factors for just COVID-19. This combined design can be applied to future epidemics, but also to study causes of non-epidemic disease.

Widespread testing is essential for monitoring the Coronavirus 2019 (COVID-19) pandemic.[1, 2] Most countries are focussing on testing persons with symptoms to identify patients with Severe Acute Respiratory Syndrome Coronavirus-2 (SARS-CoV-2) infections. Ideally, this should be coupled with random/representative population testing to follow the epidemic in the population.[3] However, there is much that can be learnt about the causes of COVID-19, even if only symptomatic people are tested. Still more may be learnt by conducting formal test-negative design studies, with additional population controls, thus yielding three linked case-control studies. In this paper, we describe how these combined study designs can enhance understanding of risk factors for symptomatic SARS-CoV-2 infection in the COVID-19 pandemic.

### *Essence of the Test-Negative Case-control Design (TND)*

Test-negative case-control studies[4-9] are based on persons who undergo testing because they present with signs and symptoms of a particular disease. The cases are those who test positive for the disease, and the controls are those who test negative - the latter will have another reason for their signs and symptoms, most likely another disease. [8] These 'cases' and 'controls' usually come from one geographical population, although not everyone in a particular area may present for testing (and some people may come from outside the area).

TNDs involve comparing the odds of a given intervention (e.g. vaccine receipt) or a given risk factor (e.g., oral contraceptives) among symptomatic persons who test positive compared to those who test negative. Given assumptions described in the literature,[8] it can produce effect estimates (odds ratios) which are generalizable to the general population (See APPENDIX A for more detail). The approach is most commonly known for its use in assessing vaccine effectiveness,[4] but has also been applied to study risk factors for antibiotic resistance,[5, 10] and to estimate risk factors in circumstances in which diagnostic bias was suspected, for example in studies on oral contraceptives and venous thrombosis, and on aspirin use and Reye syndrome.[8]

Test-negative designs allow us to obtain quick answers to important questions. Additionally, by design, they protect against some forms of bias which are otherwise difficult to control. People who are tested for a disease will not be representative of all those who have the disease (unless everyone in the population is tested) - usually, they are more likely to have severe symptoms, and more likely to seek medical help. This help-seeking behaviour is affected by many factors such as age, gender, socioeconomic status, access to health care, proximity to testing facilities, severity of symptoms, personality, and insurance coverage. In a test-negative design the same selective forces that lead individuals to be tested will operate on both those who test positive and those who test negative. There is a substantial literature on this study design,[4-9] and it is generally agreed that it can produce valid effect estimates under the assumption of similar selection pressures for the test-positives and the test-negatives.

*Reasons for considering the TND in the COVID-19 pandemic*

Insights into risk factors for COVID-19 can be gained by collecting the same information on symptomatic individuals who test positive and those who test negative, i.e., by performing a test-negative case-control study. Because the test-negatives belong to the same population (i.e. people who would come for testing if they had symptoms of COVID-19) as the test-positives, this may give timely and locally relevant insight into the causes of SARS-CoV-2 infection in different communities (urban, rural), in communities with many cases, or communities with few cases.

Direct comparisons of test-positives to test-negatives (comparison TND in Figures 1 and 2) can yield insight into specific risk factors for becoming infected and symptomatic with SARS-CoV-2: these may include age, sex, race/ethnicity, socioeconomic factors (e.g. income, education), occupational exposures (e.g. healthcare workers performing aerosol-generating procedures, delivery drivers, teachers), contact patterns (e.g. household exposure to confirmed case, crowding, travel histories, childcare responsibilities), geographic residence (e.g. urban versus rural), behavioural factors (e.g. shopping locations, smoking), medical risk factors (e.g. immunodeficiency), and genetic factors (from the swabs or blood sample taken for viral diagnosis, which will also contain human cells).

Some of this information might already be routinely collected. If a sufficient number of test sites test large numbers of people, *different* types of additional information may be asked at *different* testing sites – so as not to burden the test sites and to be able to adapt questionnaires to evolving questions. Some risk factors may be immediately important for local decisions, others more widely or more theoretically. The data can be analysed like any other case-control study, although consideration should be given to assessing possible interpretation issues arising because both the cases and controls are drawn from a subgroup of the general population (see also APPENDIX A).[8]

An interesting variant might be to study risk factors for antibody seroprevalence (instead of new infections) which is an approach that investigates cumulative risk of infection rather than incident infection. Some of its uses are discussed below.

*Critical reflections on the interpretation and feasibility of the TND in the COVID-19 pandemic*

The TND involves a comparison between persons who test positive for SARS-CoV-2 and persons who test negative but who have similar signs and symptoms. The test-negatives will have another reason for their similar signs and symptoms - most likely they will have another viral respiratory infection. Some exposures (e.g. overcrowding), will increase the risks both of COVID-19 and of other respiratory infection. Thus, the TND can only identify those risk factors that are either totally distinct or clearly different in magnitude from the risk factors of illnesses that manifest with similar symptoms. For example, if living in crowded conditions equally increases the risks both of COVID-19 and of other respiratory infections, then the proportions living in crowded conditions would be similar in the test-positives and the test-negatives. On the other hand, if male sex was a risk factor for symptomatic SARS-CoV-2 infection, but not for other respiratory infections, then more of the test-positives than the test-negatives would be male.

A second concern is about the sensitivity and specificity of the test. Although RT-PCR testing has a high specificity for SARS-CoV-2, the sensitivity can vary in relation to timing of symptom onset,[11, 12] the bodily fluid tested,[13] and the assay used.[14] There will be misclassification of cases and controls. This can be expected to be 'non-differential' (whether the test works correctly on a particular person is unrelated to exposures such as crowding). Such non-differential misclassification of exposure or disease is a known problem in case-control studies, and it usually results in bias of the effect estimate towards the null (an underestimation of effects). However, there is a major difference between the usual case-control study and the TND. In the standard population-based case-control study, the false-negatives remain part of the source population, and only a (small) fraction of them will be sampled and end up in the control group. In contrast, in the TND it is certain that at a particular test site all false-negatives will be included in the control group (the test negatives); similarly, all false-positives will be included in the case group - since all persons tested at the test site will be in the study. This may lead to stronger misclassification which has most consequences in situations wherein the proportion of COVID-19 relative to other respiratory disease among the persons tested is either very high or very low. There is an extensive literature on sensitivity analysis for standard case-control studies,[15, 16] which essentially involves making assumptions about how large this misclassification would be, and these methods can be adapted to the TND situation. See APPENDIX B for further details.

An additional reflection is about seasonality; it is not known at the time of writing how the incidence of SARS-CoV-2 infections over the calendar year will evolve. There is the possibility that other respiratory viruses such as influenza might disappear during summer,[11] whereas the SARS-CoV-2 may continue to circulate; in that situation, there may not be sufficient test-negative controls.

In the earliest phases of the COVID-19 epidemic, testing for acute infection may not have been done. The above-mentioned variant of a TND on seroprevalence of antibodies may be a solution. For example, in a hospital-based setting, serologic testing of healthcare workers for antibodies specific to SARS-CoV-2 may yield insights into exposure risks that could have been missed due to the initial lack of testing for acute infection. Of course, this will miss persons who did not survive.

As the epidemic progresses, risk factors for having had the infection might become negative risk factors for new infections. For example, bus drivers may have been frequently infected early on; if these infections conferred sufficiently strong immunity, bus drivers may turn up later in the epidemic mainly as 'test-negatives' for *acute* SARS-CoV2-infection when having signs and symptoms of respiratory disease (from another virus). In addition, measures might have been taken to shield bus drivers (passengers only entering via rear doors; obligatory unoccupied seats). While this muddles the estimates of risk factors, the immunity conferred by earlier infection as well as the preventive measures taken earlier are worthwhile goals of a TND study. This paradox requires to ideally investigate the evolution of the associations over the course of the epidemic, with background knowledge beyond the data that were obtained; as a minimum we need to look separately at the 'upward' and the 'downward' phase of the epidemic curve. To verify a potential role of immunity in an epidemic, subsamples of test-positives and test-negatives for acute infection might be tested for SARS-Cov-2 antibodies. The mentioned variant of studying antibody seroprevalence in a TND, instead of acute infections, may partially solve the problem.

***Adding standard case-control studies with a control group representing the underlying population***

As noted, a TND can potentially identify risk factors for COVID-19 that differ from those for other respiratory infections, either in kind or in magnitude, but will not identify risk factors that the test-positives and test-negatives have in common. On the other hand, comparing test-positives with general population controls will tell us about risk factors for COVID-19, but does not tell us which factor is specific for SARS-CoV-2 rather than respiratory infections in general. Thus, the ideal situation is to also have a comparison of the test-negatives with the general population. This strategy has already been applied as an extension of TNDs of antibiotic resistance.[5, 10]

Below, we outline two different strategies to obtain population controls: first the use of 'accompanying persons' (e.g., friends or household members) as 'matched controls', and second the use of a random sample from general population databases with country-wide health care and other registered information from that population. In a separate section, we will briefly discuss other possibilities for choosing controls that might be more useful in a diversity of situations. First we discuss the benefits of population controls.

***Benefits of added population controls in separate case-control studies with test-positives and test-negatives***

The importance of having population controls can be seen from Figures 1 and 2, which respectively refer to the situation with accompanying persons as controls, and to the more general situation of population controls. A comparison of the findings from the test-negative design (TND) with a case-control comparison of the test positives and their population controls (CC-POS), and a separate comparison of test-negatives with the population (CC-NEG), will enable us to assess which risk factors are specific to COVID-19 and which are risk factors for all respiratory infections (including SARS-CoV-2) in general. If these studies were all perfect, one would be able to calculate the results of any one contrast from the two others, e.g., the results of comparison CC-POS should logically follow from combining the results of the TND comparison and CC-NEG (if the odds ratio for male sex is 1.0 in the TND, but is 2.0 in comparison CC-NEG, then it also should be 2.0 in comparison CC-POS). In reality there might be differences due to sampling and/or unknown selection biases. Thus, although it would be sufficient in theory to only conduct the TND and comparison CC-POS, it remains valuable to conduct comparison CC-NEG. This enables 'triangulation' [17] with information about differences in risk factors between symptomatic test-positives and test-negatives, and two case-control studies of test-positives and test-negatives with their population controls.

***Accompanying persons as a control group***

Symptomatic persons who go for SARS-CoV-2 testing may be accompanied by other persons, e.g., household members, relatives or friends. Thus, it may be expedient to ask an accompanying person to volunteer the same information (e.g. completing a questionnaire) at the time of testing the person with symptoms – this may be done before the test result is known. These persons are members of the source population which generated the cases and should not have COVID-19 symptoms. Note that for this design it is not necessary to carry out the test on the accompanying person.

For both the test-positives (COVID-19 cases) and the test-negatives (controls with other respiratory infections), the accompanying person can be seen as a matched population control. Such approaches have been widely used in epidemiology, and the strengths and weaknesses have been extensively discussed.[18, 19] Briefly, using friends, siblings or spouses as matched population controls, has the advantage of logistic convenience, and may indirectly match for various risk factors (e.g. socioeconomic status, availability of health care, health seeking behaviour). As with any other pair-matched case-control study, this necessitates a pair-matched analysis. Essentially the matched analysis focusses on the subgroup of case-control pairs where the case and control differ with respect to the exposure under study: a pair-matched analysis is an analysis of the differences that *remain* between cases and their controls despite them being made 'more equal' by the matching.

This strategy leads to the case-control comparisons represented in Figure1: test-positives with their accompanying persons (comparison Case Control-POS), and test-negatives with their accompanying persons (comparison Case Control-NEG). Comparison CC-POS enables us to study directly the differences in risk factors between a person with COVID-19 and a control person without respiratory symptoms. Thus, in this analysis all risk factors that increase the risk of COVID-19 (some of which will also be risk factors for other respiratory diseases) will be seen to differ between cases and controls. Comparison CC-NEG enables us to directly assess risk factors for the mixture of other respiratory pathogens (e.g. influenza virus, rhinovirus) that could be causing symptoms similar to those of COVID-19.

*Critical reflections on the interpretation of case-control studies with accompanying persons as matched controls*

The use of 'friend' controls leaves the choice of the control to the case and not to the investigator [see pages 119-120 in[19]]. Friend controls may be quite similar to the cases, which is an intended benefit of matching. However, they may have some possible inherent biases [pages 119-20 in [19]], for example popularity of certain persons, or that extroverts are more often mentioned as friends. We should stress that the problem is not that the cases and controls are made too similar - this problem applies to all matched case-control studies and is addressed by taking the matching into account in the analyses.[18] Rather, the problem is that they may be made similar in ways that the investigator cannot control, and certain types of persons might be more valued to be named as friends.

A second issue is that the accompanying persons of the test positives in the CC-POS comparison may be as yet asymptomatic carriers of SARS-CoV-2. A common reflex might be to want to know this and to remove these persons from the analysis. However, apart from involving logistically difficult additional testing of the accompanying persons, it is not necessary. This is explained in detail in APPENDIX C. Briefly, the studies are based on the source population which would come for testing *if they develop symptoms*; the cases are people who have actually developed symptoms and come for testing. The controls should be a sample of the source population which generated the cases.[20, 21] Since the accompanying persons came with their index person for testing, it is reasonable to assume, that they would also have come for testing at the same facility if they had developed symptoms.

*Random sampling from country-wide general population health care and other data-bases*

In regions or countries where all health care activities are registered (prescriptions, hospitalizations, test-results, etc.) in digital databases, it may be possible to use a different type of control group, comprising a control population randomly sampled from the region or country as a whole. While analyses based on existing databases may lack the immediacy and flexibility of point-of-care data collection of persons who are tested, the advantage is that data are recorded *prospectively in past time*, and the epidemic can be analysed, and re-analysed, in its several stages (e.g., in relation to the implementation of social distancing and lock-downs).

The analysis of the COVID-19 epidemic would start with recorded data of test-positives and test-negatives for SARS-CoV-2 in the total administrative population of a country or region. While this limits information to health care data that are registered at a particular point in time, an advantage is that health care data that have been registered before (e.g., pre-existing diseases and prescriptions, prior hospitalizations etc.), can be added, as well as other data such as data on crowding, income, level of education, etc., from other data-bases.

A single control group can be used for both CC-POS and CC-NEG (see Figure 2). This allows us to randomly sample several times as many controls as there are test-positives and test-negatives combined. For efficiency purposes, the cases and controls, as well as the random population control, might be limited to an age bracket, say age 15-74, as there will be few symptomatic SARS-C0V-2 cases below age 15, and persons above age 75 may not be tested nor hospitalized. Age and sex matching are undesirable in the context of COVID-19 as these may be determinants of infection and disease course. It is always possible to stratify for age and sex, as the numbers will be sufficiently large. Matching on being alive at the index date of the cases (i.e. the date of testing) might be considered; however, this might be replaced by a control group that is composed of persons being alive in the middle of the month in which persons were tested.

It might be objected that we use 'two different control groups' for one case group, which is often frowned upon, because if the findings with the different control groups are different, the investigator has to 'choose' which one is 'right' [see pages 121-2 in [19]] However, in interpreting the combination of the TND with population control groups, neither is 'the right one', as both point to a different contrast. This can be learned from a test-negative case-control study on urinary tract infection with antibiotic resistant bacteria in contrast to infections with sensitive bacteria, with added population controls to both groups.[10] In this study, male gender proved a strong risk factor for antibiotic resistance in the test-negative design, while female gender was a strong risk factor for urinary tract infections in comparisons with the population. This seems like a 'reversing' of risk factors, but is logical because men generally only acquire urinary tract infections at older ages, subsequent to prostate or other pathology which puts them also at risk of acquiring resistant bacteria.[10] This analysis and reasoning is explained in Appendix D. It shows how the triangulation of the test-negative design with population controls leads to identifying the right causal pathways.

*Other population control groups*

Many alternatives for population control selection are possible, depending on the situation in different regions or countries. Some of these other options will be closer to the flexibility of the 'accompanying persons' control groups, others will be closer to the advantages of using existing databases. The appropriate approach will depend on local considerations. For example:

- records from General Practitioner databases (e.g., in the UK, in some regions in Italy or Spain), or third-party payers and insurers (Medicare, Medicaid, health maintenance organizations in the US), which will also allow for database-centred research.

- if patients are presenting in a special set-up organized by groups of General Practitioners (e.g., "Corona-test-sites" managed by several GP practices where patients are referred to for testing), control persons may be a sample from these general practices; this sample could be matched to the practice of the referring GPs, or weighted according to the size of the referring GP practices. This might facilitate the collection of specific new information relevant to local situations in individual practices (e.g., the use of local sports facilities).

- if patients present to outpatient clinics or hospital departments, a control group of non-respiratory out- or inpatients might be constructed to represent the catchment population of the hospital; such patient controls used to be common in pharmacoepidemiology.[22]

*Critical reflections on the choice of population controls*

It is imperative to consider to what extent the test-positives and test-negatives from a TND are representative of all cases in the general population, i.e., whether the general population can really be seen as the source population for the tested persons. There are two considerations: patient selection and doctor's preferences.

*Patient selection*

Not all diseases present equally to health care facilities. In countries with universal access to health care and relatively standardized care, it is likely that, for example, almost all solid cancers (colon, lung, etc.) with onset before age 70 will ultimately be diagnosed and recorded. That is not the case for self-limiting diseases such as influenza-like illnesses or headache. Many persons will just stay home. Only persons who worry or have more severe symptoms will present themselves to a primary care service. Still, in a country with universal access to relatively standardized care, the types of person who present themselves at several points of care will be roughly similar, and if testing is done for SARS-CoV-2, test-positives and test-negatives can be seen as drawn from the same underlying general population.

The type of person that is tested may differ between countries, however. During the initial wave of COVID-19 in February-March in Europe, testing for persons with minor symptoms was available in Germany; in the Netherlands only persons with symptoms that were sufficiently severe to be hospitalized were tested. Because these were country-wide measures, in both countries the general population may be seen as the source population. In this context, it should be noted that if (self) selection is based only on severity of disease, this will not create a bias in itself within one country.

Problems only arise if there are other factors which affect presentation for testing, given a particular level of symptoms. For example, consider private health care facilities that are only accessible to individuals who can afford them (these facilities exist in many guises: from standard private health care coming as an employee benefit, to facilities only available to the very rich). A comparison of the test-positives and test-negatives from such facilities with a general population control group may not be warranted, because of inherent differences in socio-economic status, medical care, and life style. Among persons tested in private facilities, both test positives and test negatives are, for example, unlikely to live in very overcrowded conditions. Thus, a better control group, representing their own source population, might be composed of other person (or patients) who make use of the same health care facilities. This means that one may not be able to study all of the causes of the disease (e.g. if poverty is an underlying cause, but no one who accesses these health care facilities is low-income). Also, if we suspect that there are differences in access to testing for persons with very mild symptoms or without symptoms (depending on type or health insurance or wealth), but few such differences for severe disease, then one might restrict the analysis to the subgroup of tested individuals with more severe symptomatic disease.

*Doctors' preferences*

Even in settings with relatively standardized care, there might be variation in testing strategies for a new disease like COVID-19 between (primary care) practices. The existence of physician preference has been studied in different countries.[23] Physician preference can be based on a different interpretation of the literature on topics where there is not yet consensus, or on implicit biases (regarding age, sex, ethnicity). If this is suspected to have been the case, it might be better to select population controls from the GP practices of the individuals who underwent testing – and to approach this in the analysis as a form of 'matching'. Matching by GP practice would, however, limit the ability to compare between catchment areas and lead to loss of information about regional differences. Once again, restricting the analysis to patients with severe disease (whether test-positive or test-negative), may suffer less from self-selection and testing preference by doctors. An analysis according to severity can also be added as a sensitivity analysis.

## Discussion

An ideal approach for identifying risk factors for COVID-19 would involve random/representative population sampling.[3] However, in the surveillance efforts that are being developed when an epidemic is unfolding, population-based testing often is limited by laboratory capacity (i.e., due to unprecedented demands for reagents and trained technicians), funding and political will. The first thoughts of decision makers are to facilitate testing for people with symptoms who became ill recently, either in order to isolate, or in order to know which treatment trajectory is necessary if symptoms worsen.[1, 2]

The situation with COVID-19 remains urgent in many parts of the world, and it is important that the best possible use is made of information collected in the process of widespread testing of symptomatic persons. Therefore, there are research and public health benefits in employing a test-negative case-control design, combined with case-control studies with population controls added to it. Still, such collection of information has to be as 'light' as possible, in order not to disturb the primary medical aim: to test people for their own benefit and for controlling the epidemic. The proposed data collection can be done with minimal extra effort, it would roll along with the epidemic, and can potentially yield important information at much less cost, and with greater ease, than doing genuinely random population repeated sampling and testing. In situations where extensive databases exist, data will have been collected as the epidemic unfolded, and then kept frozen in time in the databases. This allows investigators to return to the data and evaluate the course of the epidemic with new hypotheses. At the time of this writing, several efforts are underway to set up collections of types of questions and data of interest to study the COVID-19 pandemic.[24, 25]

Adding general population controls yields linked case-control studies (the TND, CC-POS and CC-NEG) and creates a triangulation situation[17] for inferences about local as well as general factors that drive the pandemic. Follow-up of test-positives or test-negatives or other 'add-ons' will lead to better understanding of the course of the disease. In particular, a follow-up starting from a test-negative design of acute disease may be a good starting point to provide information on the degree and duration of protection from (re)infection conferred by having had a SARS-CoV-2 infection previously and/or specific antibody titres, because such studies might need exact matching on date and place of testing.[26] Finally, having an infrastructure for a test-negative design already established in different settings may be a valuable base to evaluate the effectiveness of interventions such vaccines when they become available, or other measures to limit transmission.

The combined test-negative and population-based case-control design may be useful, not only for recurrent waves of this epidemic or other epidemics, but also to study causes of non-epidemic diseases, in circumstances where a test-negative design would have advantages but in which its inferences could be strengthened by comparing with population controls.

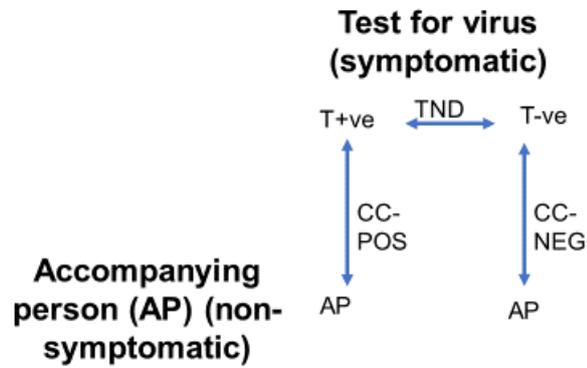

**Figure 1:** Test-negative design and case-control studies with accompanying persons

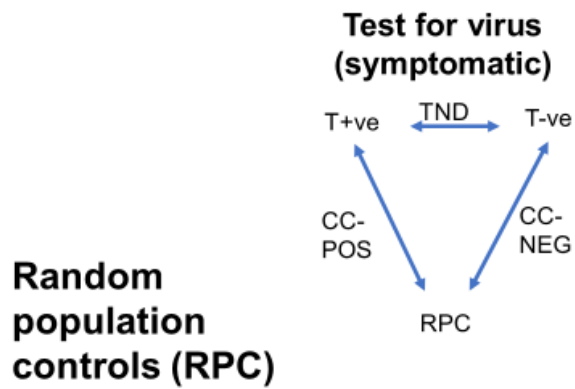

**Figure 2:** Test-negative design and case-control studies with random population controls

# APPENDICES TO: Test-Negative Design and Case-Control studies with population controls during widespread testing of symptomatic persons for the presence of SARS-CoV-2

Jan P Vandenbroucke[1,2,3], Elizabeth B Brickley[1],

Christina M.J.E. Vandenbroucke-Grauls[4], Neil Pearce[1]

(1) Departments of Medical Statistics, Non-communicable Disease Epidemiology and Infectious Disease Epidemiology, London School of Hygiene and Tropical Medicine, London, UK

(2) Department of Clinical Epidemiology, Leiden University Medical Center, The Netherlands

(3) Department of Clinical Epidemiology, Aarhus University, Denmark

(4) Department of Medical Microbiology and Infection Prevention, Amsterdam UMC, The Netherlands

Corresponding Author:

Jan P Vandenbroucke, Leiden University Medical Center, Dept. Clinical Epidemiology, PO Box 9600, 2300 RC Leiden, The Netherlands, E-mail: J.P.Vandenbroucke@lumc.nl,

**Appendix A: Assumptions that underly generalizability of estimates of a Test-Negative Designs to a broader population**

The TND design has been described in detail in numerous papers. In a recent paper two of the authors of the present paper presented this study design as a variant of 'case-control studies with other disease controls'. [1] These include hospital-based case -control studies in which the cases are identified through a particular health-care facility, and the controls are other patients at the same health-care facility.

Case-control studies with diseased controls have a long history in epidemiology, from the first case-control studies of smoking and lung cancer (where lung cancer patients were compared to other patients of the same wards), to several applications in pharmacoepidemiology,[2] and applications in cancer registries.[3] Such studies are recognized as having produced some major and valid findings, such as the association between smoking and lung cancer.[4] However, they rest upon the assumption that the main risk factor under study does not cause the 'other diseases' which are used as controls.

The main difference between the TND in general and other case-control studies with diseased controls is that in the TND the cases and controls are persons with similar signs and symptoms, but who 'test' differently on a crucial test for a particular disease. As such, the controls (those who test negative) will usually have another disease with similar signs and symptoms to the disease under study.

This means that in the particular context of the application of the TND to risk factors for COVID-19, the standard assumption of the hospital-based case-control study, that the exposure of interest is not a cause of the other disease, is not true for many exposures. The other diseases will mainly be other respiratory infections, and several of their risk factors will be similar to those for COVID-19. As mentioned in the paper, the TND will not detect risk factors that are of equal strength for COVID-19 and the diseases of the controls; however, it will detect differences in magnitude of a particular risk factor (say, crowding), or risk factors that would be totally specific for COVID-19. This situation is similar to the situation of another TND on urinary tract infection with antibiotic resistant bacteria (Extended Spectrum Beta Lactamase producing bacteria) in contrast to sensitive bacteria, with added population controls to both groups.[5] In that study, recent hospitalizations proved to be a risk factor that was specific for acquiring infections with resistant bacteria; other findings could only be interpreted by combining the TND with the population controls (e.g., the effect of maleness which was a strong positive risk factor in the TND analysis, but the inverse was true in the comparison with the population).

The basic question is whether the OR generated by a TND, for a factor that is a risk factor for the disease of interest and not for the control diseases, can in principle be the same as the OR that would have been generated by a population-based case-control design. This question has either been answered with strong doubts by some[6] because the TND does not have a well-defined underlying source population; others have answered it positively, some with a few qualifications;[7-9] some with more qualifications.[10] In essence, we agree with the affirmative answer to that question because the TND permits a positive answer to one 'sentinel question' about the validity of a control group: would this person, who does not have the disease, have presented him/herself for testing in that same facility if s/he would have had the disease under study.[1] The answer for a TND is obviously positive. As a minimum, the results of the TND are valid for the population of 'the tested'.

An additional subtlety is that in a sense, one might even state that in exceptional circumstances where the inclination to be tested is highly linked to characteristics of the tested person, or to suspicions or habits of doctors who serve a particular segment of the population, the standard case-control study with 'test-positive cases' and random general population controls could produce wrong answers, since the type of persons going for testing at particular medical facilities might be self-selected in a way that the random sample of the underlying population is not. [If they were billionaires with doctors catering for that category of persons, there would be habits that would not differ from other billionaires but might differ quite a lot from the general population]. The TND remedies that situation by its design by choosing controls with the same selection pressures, and will in such circumstances produce a valid answer for the tested.

The next question is whether the OR for the TND (= those who go for testing) is applicable to the general population from which it is drawn, or for other populations, again for risk factors that are not known risk factors for the control diseases. It is possible that the TND may deliver unbiased odds ratios for the source population of the 'tested' (i.e. everyone who would have been tested if they had developed symptoms), but these findings may not be completely 'transportable' to the wider general population or other populations – for example, the effect might be higher among the tested than among the general population. However, there would only be lack of generalizability if there were strong effect modification, and the distribution of effect modifiers was markedly different in other populations.[11]

For example, is it possible that a genetic a risk factor for Covid-19 is found in a TND study, but not in the wider population? This seems unlikely. More importantly, these issues of generalizability to other populations is an issue in many case-control studies, and is not unique to the TND. For example, one might ask whether the original studies on smoking and lung cancer in (white Anglo-Saxon) male British doctors in the 1960s would be equally applicable to lawyers, actors, sewage cleaners, or to woman and non-white non-Anglo-Saxon persons in general. Issues of generalization apply to all case-control study findings, not just the TND.

**APPENDIX B. Sensitivity analysis: quantification of misclassification bias in TND due to false test-negatives and false test-positives**

If one assumes particular values of Sn, Sp in a particular study, then one can estimate the bias that is introduced by misclassification of the disease. This can be used to do a sensitivity analysis where we assume a range of possible values of Sn, and Sp. In theory, one can then do an adjustment. However, given the uncertainties involved, such corrections are usually conducted as additions to the main analysis (which would report the unadjusted associations as the main results but also mention the sensitivity analysis findings), rather than substitutes for it.

This Appendix will explain first the situation of disease misclassification in a TND study that is intended to contrast COVID-19 patients vs. patients with other respiratory disorders (mostly other viral infections); second it will explain how different proportions of COVID-19 vs. other respiratory disorders will influence the role of sensitivity and specificity; and thirdly, that it is possible to calculate back from an observed 2x2 table to the 'true' table, given assumptions on sensitivity and specificity.

One may start from the assumption that the specificity of the RT-PCR is very high, except for mishandling of specimens, mishandling of test information (mixing up of persons), or batch contamination. In contrast, the sensitivity may depend on several factors such as the severity and duration of the disease, the way the sample was obtained, as well as the performance of the test itself (See paper).

To see what might happen in the TND situation, a simple numerical example is useful. All persons with COVID-19 with a number of characteristics (a certain degree of symptoms, and a certain degree of worry, personality, access to health care etc.), will have presented themselves to a particular test facility. If only 70% of them will test positive (sensitivity is 70%), the remaining 30% will be false-negatives, but as they have all presented themselves, they will be classified as test-negatives along with the group of true test-negatives, i.e. persons with other respiratory diseases. If we assume that the number of false-positives is negligible, and suppose that at a particular test-site there are 70 test-positives and 150 test negatives, then the true number of COVID-19 patients will be 70/0.7 = 100. This means that 30 of the 150 test-negatives are actually false-negatives, which is 20%. There will thus be 100 COVID-19 patients out of 220 tested. Thus, all ORs on the observed numbers will be biased towards the null. In principle, we can recalculate the true OR without test misclassification by a standard procedure that surmises that the false-negatives amongst the test-negatives should have the same exposure frequencies as the test-positives (i.e. misclassification is non-differential).[1]

In this example, we have not yet taken into account that a (probably small) number of the test-positives are actually false-positives, which in this numerical example is negligible, but that is not always the case, as this depends on the proportion of true COVID-19 patients among the persons tested.

---

[1] Suppose that 42 of test-positives were men (60%), and only 50% of test negatives were men, this yields an OR for maleness of 1.5. The real percentage of males amongst the true test-negatives can be calculated as 'X' in the equation: 0.50 = (0.8 x X) + (0.2 x 0.60), which gives an X of 0.475. Then the true odds ratio becomes: (0.60 x 0.525)/(0.475 x 0.40) = 1.66 (rounded off).

To demonstrates what happens with mutual misclassification, in different scenarios of prevalence of COVID-19 relative to prevalence of other respiratory disorders amongst the persons tested, we can make Table 1, following a reasoning similar to that by Flegal et al.[12] From Table 1 it becomes apparent that if the prevalence of current COVID-19 infection is high relative to the prevalence of other respiratory viral diseases amongst the tested (say, in summer months at the peak of an epidemic in a population without immunity), the main driver of the misclassification is the sensitivity of the test: a large proportion of test-negatives might actually be false negatives. Inversely, in the situation of an almost disappearing epidemic in winter months (say, when a vaccine would be available and it would be winter), the number of test-positives that would actually be false-positives will become quite large, i.e., the specificity becomes more important. In each of these extreme situations, there is misclassification, but it is less severe in the latter instances (sporadic COVID-19).

It is possible to make approximate assumptions about the ratio of the estimated OR to the true OR, based on tables like this, with appropriate estimates of sensitivity and specificity and relative prevalences of non-Covid-19 respiratory diseases amongst the tested.

**TABLE 1: Variation of observed OR for different True COVID-19 percentages amongst the tested with the same sensitivity, specificity, and the same exposure frequency in true test-negatives (assumed to be 50%), and same true underlying OR of 2.0.**

| Fraction TRUE COVID-19 | SENS | SPEC | TRUE Odds Ratio (rounded) | **OBSERVED Odds Ratio (rounded)** |
|---|---|---|---|---|
| **0,97** | 0,70 | 0,99 | 2,00 | **1,07** |
| **0,90** | 0,70 | 0,99 | 2,00 | **1,21** |
| **0,50** | 0,70 | 0,99 | 2,00 | **1,69** |
| **0,10** | 0,70 | 0,99 | 2,00 | **1,80** |
| **0,03** | 0,70 | 0,99 | 2,00 | **1,58** |

Finally, there is the possibility to calculate back from an observed 2x2 table to the "true" 2x2 table, given assumptions on sensitivity and specificity. See Chapter 6 on "Outcome misclassification in Lash et al.[13] This "correction", is based on a single observed 2x2 table, when there has been confounder adjustment, the adjusted odds ratio will differ from the crude odds ratio, and therefore one cannot use the crude 2x2 table for the "correction" – more complex methods are available if one wishes to also correct for confounders while "correcting" for misclassification.[14]

**APPENDIX C: Reason for not omitting controls that are asymptomatic carriers of SARS-CoV-2**

This reasons for not omitting controls that are asymptomatic carriers of SARS-CoV-2 can be illustrated by considering a hypothetical study. Suppose we could identify the population (a subgroup of the general population) which would come for testing if they had symptoms. Ideally, one would then test all of this population and we would estimate the risk of infection in this population, and in various subgroups. In the population (P) there might be a certain number of people (T) who tested positive. Note that the population denominator (P) includes both people who currently have symptomatic infections and people who don't – it is just the total population 'at risk'. The risk of having a symptomatic infection is then T/P. If we compare two subgroups who are exposed or not-exposed to a particular factor, their risks might be $T_1/P_1$ and $T_0/P_0$ respectively, and the risk ratio (the ratio of these two proportions) would indicate the relative risk of symptomatic Covid-19 infection in the exposed and non-exposed (e.g. if the RR was 2.0 then the exposed would be twice as likely as the non-exposed to have symptomatic Covid-19 infection).

A case-control study involves studying all of the 'cases' (i.e. T) and a sample of the population which generated them (P). Thus, provided that the controls are a representative sample of the source population P (i.e. everyone who would have come for testing if they had symptoms – which is a reasonable assumption to make since they came with someone who was being tested), then the odds ratio for Covid-19 infection in the case-control study will estimate the risk ratio in the population (P) which generated the cases (T). Of course, a small number of the controls may have asymptomatic Covid-19 infection, and would have tested positive if they had been tested. But this is not a problem – they would have been part of the denominator (P) if a full population survey had been conducted, and they are therefore eligible to be selected as controls. This is analogous to the case-cohort (case-base) design which is a commonly used design for case-control studies .[15, 16]

**APPENDIX D. Differentiation of risk factors between TND and added case-control studies by triangulation**

This is a worked out example from a paper about urinary tract infections with resistant vs. sensitive bacteria.[5] In this example, the TND analysis is a contrast between patients with infections with resistant bacteria *(= test-positives)* vs. infections with sensitive bacteria *(= test-negatives).* For each group, the test-positives and the test-negatives, there is a comparison with the general population.

The TND analysis as well as the comparison between the test-positives and the population is demonstrated on partial data from the original paper which are given below as Appendix D Table 1. The other comparison, between test-negatives and the general population was in the Appendix of the original paper, and is also partially given below at the right hand side of Appendix D Table 2. Male sex is a strong risk factor for infections with antibiotic resistant bacteria in the TND analysis, but completely the reverse happens in the comparison with the general population. This makes sense: in the general population women have much more urinary tract infections than men, but men when they have urinary tract infections are generally older patients (for example, patients with prostatic disease who go in- and out of hospitals and acquire resistant bacteria in hospitals). In the data on the other comparison, between test-negatives and the general population, the OR for being a woman (the reverse of the OR for begin a man) is even more extreme, which is logical because the test-positives are removed from that comparison. If one multiplies the OR for sex of the TND with the population comparison of the test-negatives, one arrives in the direction of the population comparison with the test-positives.

**Appendix D Table 1**: Partial table of risk factors for urinary tract infection (UTI) caused by extended-spectrum beta-lactamase-producing (ESBL) *E. coli* compared with non-ESBL *E. coli* UTI controls *(a test-negative design comparison)* and population controls *(case-control comparison of test-positives to general population)*

| Characteristic | OR for ESBL *E. coli* UTI versus non-ESBL *E. coli* UTI (95% CI) | | OR for ESBL *E. coli* UTI vs. population controls (95% CI) | |
|---|---|---|---|---|
| | Adjusted for age, sex, and comorbidity | Multivariate adjusted | Adjusted for age, sex, and comorbidity | Multivariate adjusted* |
| Demographic characteristics | | | | |
| 10 years increase in age | 0.90 (0.85 - 0.96) | 0.92 (0.86 - 0.99) | 1.25 (1.16 - 1.35) | 1.16 (1.06 - 1.27) |
| Male sex | 1.55 (1.20 - 2.02) | 1.62 (1.22 - 2.14) | 0.29 (0.23 - 0.38) | 0.48 (0.34 - 0.67) |
| Citizenship: Northern Europe vs. other countries | 0.37 (0.23 - 0.60) | 0.40 (0.24 - 0.66) | 0.54 (0.33 - 0.87) | 0.39 (0.22 - 0.68) |
| Marital status | | | | |
|   Divorced or widowed vs. married | 1.04 (0.79 - 1.37) | -- | 1.18 (0.87 - 1.59) | -- |
|   Never married vs. married | 0.82 (0.57 - 1.19) | -- | 1.36 (0.95 - 1.96) | -- |
| Prior urinary tract infection | | | | |
|   Within 31-180 days before index date | 1.27 (1.01 - 1.61) | -- | 19.1 (14.3 - 25.4) | -- |
|   Within 181-365 days before index date | 1.19 (0.94 - 1.50) | -- | 10.6 (7.98 - 14.0) | -- |

* Adjusted for more factors than shown here; see original publication.[5]

**Appendix D Table 2.** Partial table of risk factors for *E. coli* urinary tract infection (UTI) with and without ESBL compared with population controls *(two comparisons of test-positives and test-negatives with general population; first comparison same as in left hand side of Appendix D Table 1)*

| Characteristic | OR for ESBL *E. coli* UTI vs. population controls (95% CI) | | OR for Non-ESBL *E. coli* UTI vs. population controls (95% CI) | |
|---|---|---|---|---|
| | Adjusted for age, sex, and comorbidity | Multivariate adjusted* | Adjusted for age, sex, and comorbidity | Multivariate adjusted* |
| Demographic characteristics | | | | |
| 10 years increase in age | 1.25 (1.16 - 1.35) | 1.16 (1.06 - 1.27) | 1.34 (1.30 - 1.38) | 1.24 (1.19 - 1.28) |
| Male sex | 0.29 (0.23 - 0.38) | 0.48 (0.34 - 0.67) | 0.19 (0.17 - 0.21) | 0.23 (0.20 - 0.27) |
| Citizenship: Northern Europe vs. other countries | 0.54 (0.33 - 0.87) | 0.39 (0.22 - 0.68) | 1.45 (1.10 - 1.93) | 1.21 (0.90 - 1.64) |
| Marital status | | | | |
|   Divorced or widowed vs. married | 1.18 (0.87 - 1.59) | -- | 1.14 (0.98 - 1.32) | -- |
|   Never married vs. married | 1.36 (0.95 - 1.96) | -- | 1.59 (1.35 - 1.87) | -- |
| Prior urinary tract infection | | | | |
|   Within 31-180 days before index date | 19.1 (14.3 - 25.4) | -- | 15.0 (12.3 - 18.2) | -- |
|   Within 181-365 days before index date | 10.6 (7.98 - 14.0) | -- | 8.90 (7.37 - 10.7) | -- |

* Adjusted for more factors than shown here; see original publication.[5]

**REFERENCES TO APPENDICES**